\documentclass{PoS}
\usepackage{graphicx,epsfig,amsfonts}
\title{Selected Results from STAR Beam Energy Scan Program}

\ShortTitle{STAR Beam Energy Scan Program }

\author{\speaker{Michal \v{S}umbera} for the STAR Collboration\\
       Nuclear Physics Institute ASCR, 250 68 \v{R}e\v{z}, Czech Republic\\
        E-mail: \email{sumbera@ujf.cas.cz}}

\abstract{Results from the Beam Energy Scan (BES) program conducted by STAR experiment at RHIC are presented.  The data from Phase-I of the BES program collected in  Au+Au collisions at center-of-mass energies ($\sqrt{s_{NN}}$) of  7.7, 11.5, 19.6, 27, and 39 GeV cover a wide range of baryon chemical potential $\mu_{B}$ (100--400 MeV) in the QCD phase diagram.  Several STAR results from the BES Phase-I related to ``turn-off'' of strongly interacting quark--gluon plasma (sQGP) signatures and signals of QCD phase boundary are reported. In addition to this, an outlook is presented for the future BES Phase-II program and a possible fixed target program at STAR.}

\FullConference{The European Physical Society Conference on High Energy Physics \\
                 18-24 July, 2013\\
                 Stockholm, Sweden}

\begin{document}

\section{Introduction}
At sufficiently high temperature $T$ or baryon chemical potential $\mu_{B}$ QCD predicts a phase transition from hadrons to the plasma of its fundamental constituents -- quarks and gluons~\cite{lattice1}. Search for and understanding of the nature of this transition has been a long-standing challenge to high-energy nuclear and particle physics community. In 2005, just five years after start up of RHIC, the first convincing arguments on the existence of de-confined strongly interacting partonic matter with unexpected properties of   perfect quark-gluon liquid, constituent number scaling, jet quenching and heavy-quark suppression  were revealed~\cite{qgp}.  A central goal now is to map out as much of the QCD phase diagram in $T$, $\mu_{B}$  plane as possible, trying to understand various ways in which the hadron-to-sQGP transition may occur.  The first STAR proposal for the BES program was published in 2008~\cite{ref_bes}. The goal is to search for the ``turn-off'' of sQGP signatures, signals of QCD phase boundary and existence of a critical point in the QCD phase diagram.  
 
After few small-statistic Au+Au exploration runs at $\sqrt{s_{NN}}$ =22 GeV  in 2005 and  at $\sqrt{s_{NN}}$ = 9.2 GeV in 2008 \cite{9gev}, STAR collected large-statistics  data sets at  $\sqrt{s_{NN}}$ = 7.7, 11.5 and 39 GeV in 2010, and 19.6 and 27 GeV in 2011~\cite{lokesh}.  
%It is noteworthy that $\sqrt{s_{NN}}$ = 7.7 GeV, which is much below the RHIC design injection energy of 19.6 GeV, remains so far also the lowest energy achieved with hadron collider.   
The almost uniform acceptance for different identified particles and collision energies at midrapidity is an important advantage of the STAR detector for the BES program. The main tracking device -- the Time Projection Chamber (TPC) -- covering  2$\pi$ in azimuth ($\phi$) and $-$1 to 1 in pseudorapidity ($\eta$) provides momentum measurements as well as particle identification (PID) of charged particles.
For the higher transverse momentum ($p_{T}$) region, the Time Of Flight (TOF) detector is quite effective in distinguishing between different particle types. Particles are identified using the ionization energy loss in TPC and time-of-flight information from TOF~\cite{tpc_tof}. The centrality selection in STAR is done using the uncorrected charge particle multiplicity measured in
the TPC within $|\eta|<$ 0.5~\cite{9gev}. 

%\vspace{-.3cm}
\section{Suppression of high-$p_{T}$ hadrons}
Study of energy-dependence  of high-$p_{T}$ hadron suppression  \cite{qgp, STAR_quench} provides a promising avenue for ``turn-off'' of sQGP measurement.  The ratio of transverse momentum differentiated spectra from central over peripheral collisions and scaled by the mean number of binary $p$+$p$-like collisions in each event is called the nuclear modification factor and is denoted by $R_{ \mathrm{CP}}$.  In the presence of sQGP, high-$p_{ \mathrm{T}}$ particles are quenched, transferring energy to lower momentum particles, causing $R_{ \mathrm{CP}}$ to be less than unity at high $p_{ \mathrm{T}}$ \cite{qgp, STAR_quench}.   Quenching competes with radial flow or the Cronin Effect or any other effects that would cause enhancement.  Later has been observed by STAR in asymmetric $d$+$Au$ collisions at $\sqrt{s_{_{ \mathrm{NN}}}}$ = 200 GeV \cite{STAR_enhanc}.  

%Moreover, the $R_{ \mathrm{CP}}$ spectra can be also influenced by the spectators, non-interacting nucleons in the collision system, so that a peripheral collision does not equate directly to a $p$+$p$ collision due to cold nuclear matter (CNM) effects.  The goal of this analysis is to determine at what beam energy suppression turns off, and to begin disentangling the causes and relative effects of quenching and enhancement.

\begin{figure}
\begin{center}
 \includegraphics[width=1.05\linewidth]{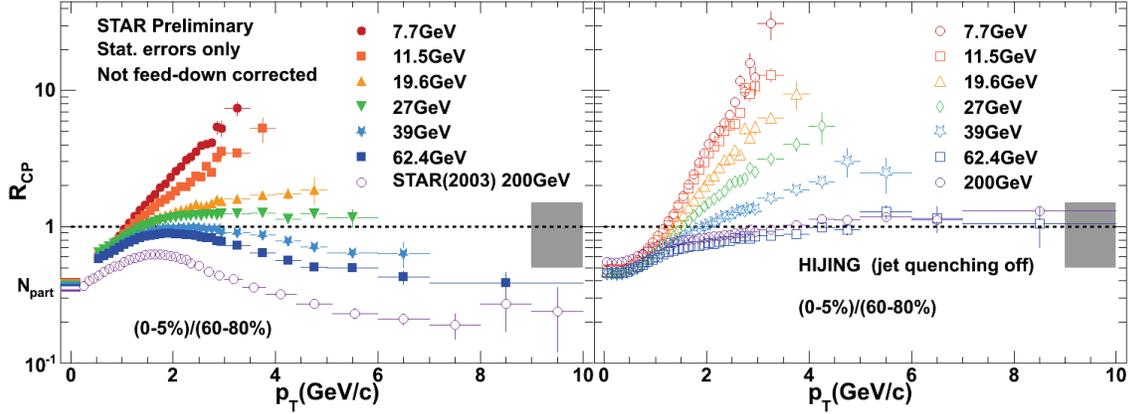}
 \end{center} 
  \vspace{-.5cm}
  \caption{\label{Fig1}{(Color online) Efficiency corrected charged hadron $R_{ \mathrm{CP}}$ (left) for RHIC BES energies.  $p_{ \mathrm{T}}$ dependent errors are statistical only.  The gray error band corresponds to the $p_{ \mathrm{T}}$ independent uncertainty in $N_{ \mathrm{coll}}$ scaling.  Charged $R_{ \mathrm{CP}}$ from HIJING simulated events (right) with jet quenching turned off.}
}
\label{RCP}
 \end{figure}

Trigger efficiency, tracking efficiency, and acceptance but not feed-down corrected charged-hadron spectra from mid-rapidity $Au$+$Au$ collisions at $\sqrt{s_{_{ \mathrm{NN}}}}$ =  7.7, 11.5, 19.6, 27, 39, and 62.4 GeV were produced for 0-5\% central and 60-80\% peripheral collisions in the STAR detector.  The spectra were scaled by binary collisions with the scale factors obtained from a Monte Carlo Glauber model \cite{glauber}.  Taking the ratio of these scaled spectra for each energy gives the $R_{ \mathrm{CP}}(\sqrt{s_{_{ \mathrm{NN}}}},p_{ \mathrm{T}})$ shown in Fig. \ref{RCP} (left) along with STAR's published 200 GeV result \cite{STAR_quench}.   The global systematic uncertainty is dominated by the uncertainty in the centrality selection for the peripheral bins, which is used in the Glauber calculation and presents as an uncertainty on the binary collision scale factor.  The same methods that produced the lower beam energy results were used to produce a 200 GeV result from 2010 data.  This measurement disagreed with the feed-down corrected result from 2003 ({Fig.\ref{Fig1} left) by 20\% and so this was folded into the overall systematic uncertainty of the other results (Fig. \ref{RCP} gray box).  The cause of this discrepancy is under investigation. For more details see \cite{Horvat:2013fla}.

%  The efficiency correction is based on single particle embedding in the 39 GeV data set for $\pi^{\pm}$, $K^{\pm}$, and $p^{\pm}$ separately which were than combined and weighted by their relative yields for a charged hadron efficiency.  The efficiency correction was extrapolated to the other data sets by making the assumption that the efficiency is the same for each particle at the same $p_{ \mathrm{T}}$ and from the same multiplicity bin.  This assumption was tested by producing the efficiencies from two data sets, 39 GeV from 2010 and 27 GeV from 2011, and ensuring that the predicted efficiency from the 39 GeV data set matched the 27 GeV efficiency.  Then differences in acceptance due to detector performance between beam energies were accounted for by using stable portions of the detector as a reference.  

Right panel of Fig. \ref{RCP} shows $R_{ \mathrm{CP}}(\sqrt{s_{_{ \mathrm{NN}}}},p_{ \mathrm{T}})$ obtained from  simulations using HIJING 1.35 model \cite{hijing} with jet quenching turned off.  The motivation for this study is the expectation that at sufficiently low beam energies, where medium-induced jet quenching has minimal effect, there would be a quantitative agreement between the charged hadron $R_{ \mathrm{CP}}$ from simulation and data.   Again, $R_{ \mathrm{CP}}$ at lower beam energies is enhanced, although we do not see a quantitative agreement with charged hadron $R_{ \mathrm{CP}}$ from data.  We do not see suppression at higher collision energies, as expected since quenching was turned off.  By running HIJING with jet quenching on and off and comparing with other models we hope to disentangle the relative contributions of jet quenching, cold nuclear matter effects, and possible contributions from radial flow or final state scattering.  The results in Fig. \ref{RCP} (left) are consistent with suppression for 
$\sqrt{s_{_{ \mathrm{NN}}}}$ $\geq$ 39 GeV.  This does not preclude medium induced energy losses at lower energies since other effects could be overwhelming this signature.  

%If this would be the case  then the energies where the deviation becomes noticeable could be considered candidates for the beam energies where a QGP is formed.   
%The advantage for using charged hadron $R_{ \mathrm{CP}}$ is that you can measure spectra to higher transverse momenta that you could not reach with particle identification.  It was also considered that plotting $R_{ \mathrm{CP}}$ vs. $x_{ \mathrm{T}}$ rather than $p_{ \mathrm{T}}$ might reveal trends in the data that were independent of collision energy.  $x_{ \mathrm{T}}$ is defined as $x_{ \mathrm{T}} = 2*p_{ \mathrm{T}}/\sqrt{s_{_{ \mathrm{NN}}}}$.  This sort of scaling was applied to spectra previously [8] where it revealed $\sqrt{s_{_{ \mathrm{NN}}}}$ independent trends at high $p_{ \mathrm{T}}$.  Such a scaling is shown in Fig. 2, using the data from Fig. 1,  and does not reveal any such trends. 

Particle yields were extracted from a simultaneous fit to $dE/dx$ distributions measured in the TPC and time of flight distributions measured in the TOF detector for each centrality and $p_{ \mathrm{T}}$ bin at each beam energy.  
%The functions used to extrapolate fit parameters for particle identification were varied in order to obtain the systematic errors for the high $p_{ \mathrm{T}}$ bins.  Efficiency corrections were obtained through track embedding and have a 5\% systematic error associated with them.  
The result (Fig.~\ref{RCPpm}, left panel) is qualitatively consistent with published data  \cite{STAR_pid_quench} in that pions are less enhanced than protons, suggesting that pions may serve as a better gauge for jet quenching within the $p_{ \mathrm{T}}$ range available through particle identification.  Considering 2.5 GeV/c \textless\ $p_{ \mathrm{T}}$ \textless\ 4 GeV/c  hadrons, the results from Fig. \ref{RCPpm} show that protons are not suppressed at any beam energy and pions go from being suppressed at higher beam energies to being enhanced at lower beam energies with a transition near $\sqrt{s_{_{ \mathrm{NN}}}}$ = 27 GeV. The result on anti-particles (Fig.~\ref{RCPpm}, right panel) is consisted with positively charged hadrons except of suppression of low momentum anti-protons observed at lower energies. Later is likely to come from large annihilation cross section of anti-protons in the hadronic medium.

\begin{figure}
 \includegraphics[width=.5\linewidth ,height =.3\linewidth]{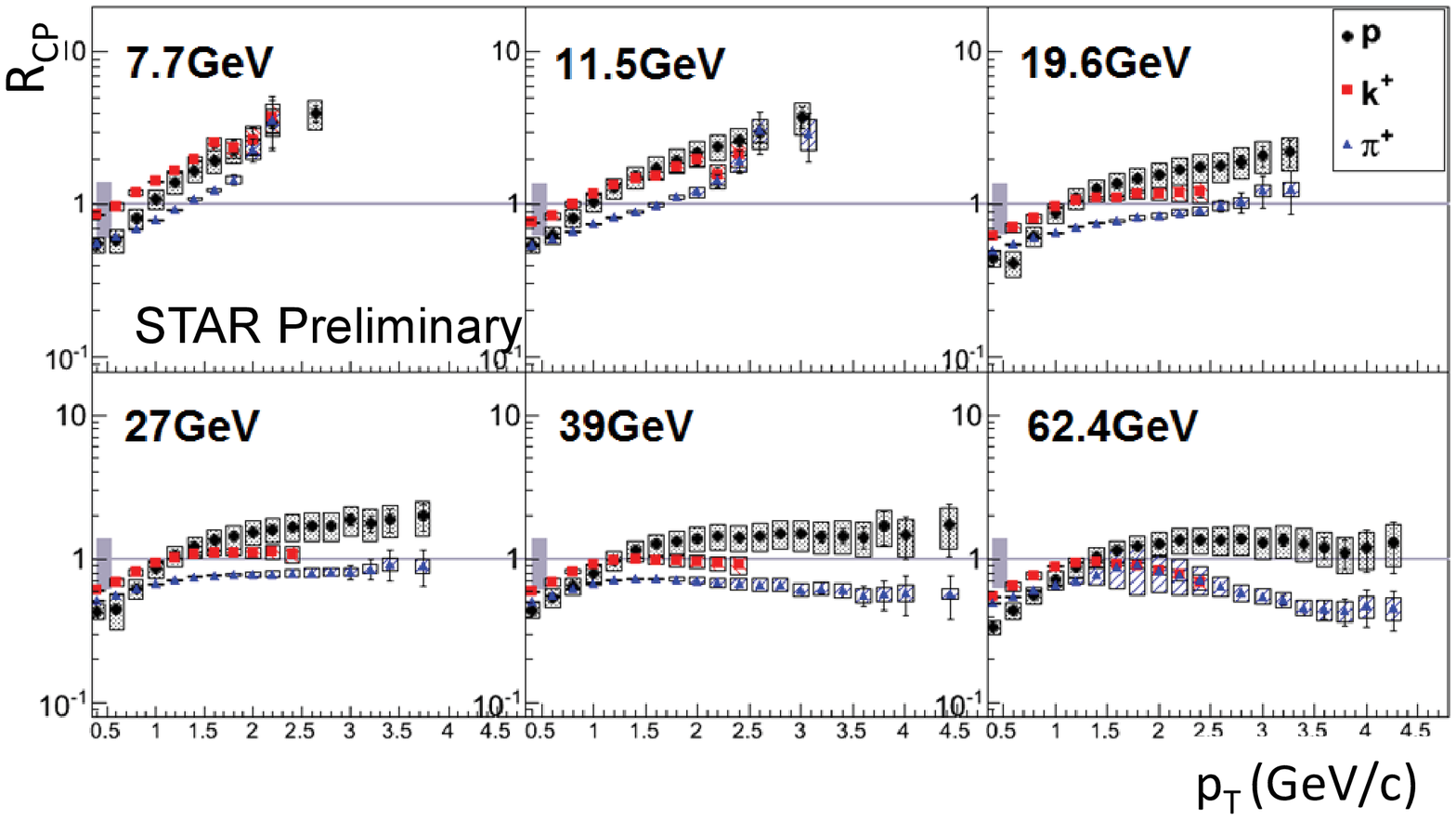} 
 \includegraphics[width=.5\linewidth ,height =.3\linewidth]{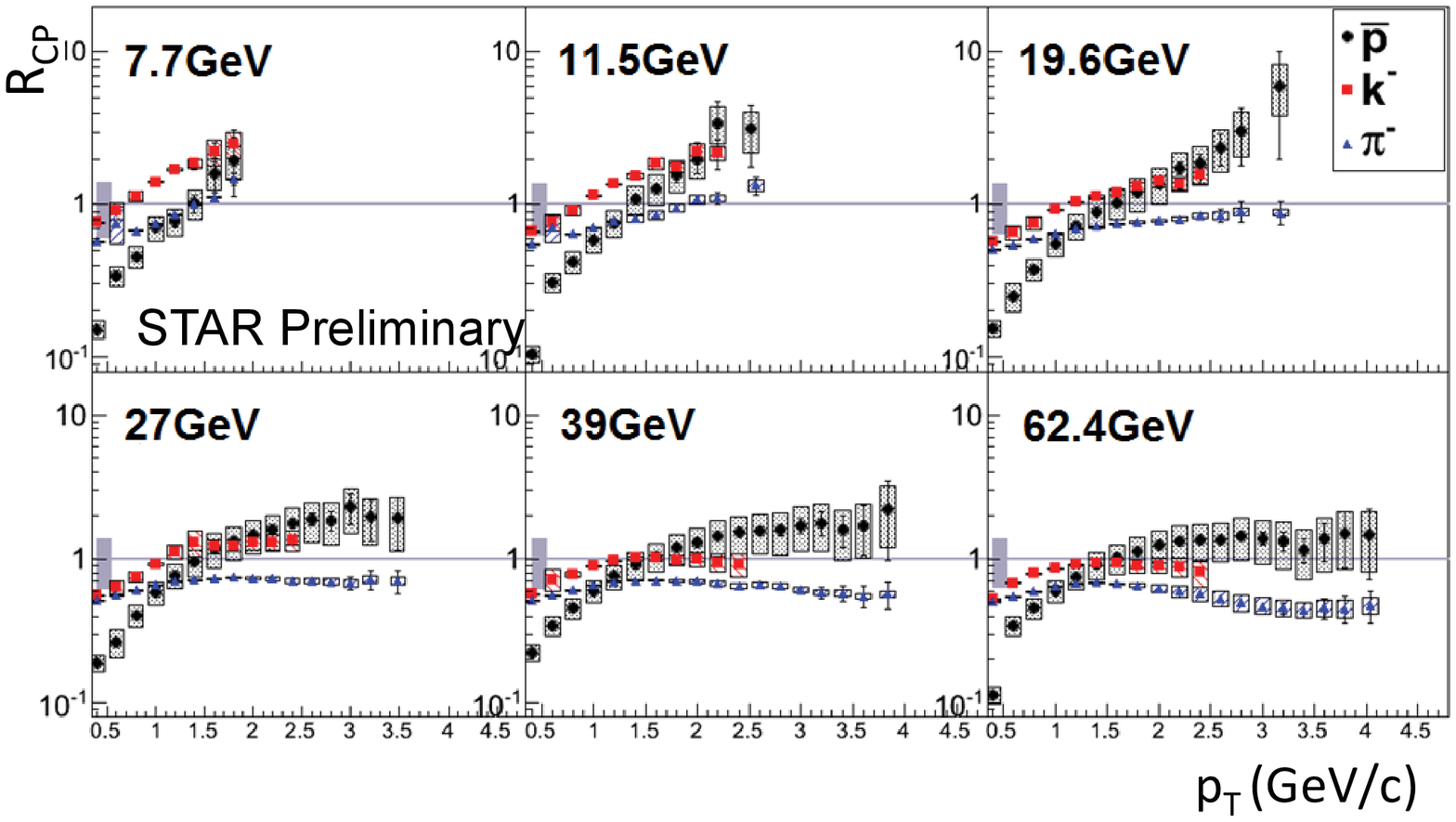}
  \vspace{-.5cm}
  \caption{(Color online) Efficiency corrected $R_{ \mathrm{CP}}$ of positive (left) and negative (right) identified hadrons for RHIC BES energies. }
 \label{RCPpm}
\end{figure}

%\vspace{-.3cm}
\section{Anisotropic Flow}
Azimuthal anisotropies $v_{1}$ and $v_{2}$ of particles with respect to event reaction plane offer insight into the equation of state of the produced matter \cite{qgp}.
The first quantity $v_{1}$ measures directed flow  due to the sideward motion of the particles within the reaction plane. It is sensitive to baryon transport, space-momentum correlations and sQGP formation.  The second quantity $v_{2}$  results from the interaction among produced particles and is directly related to transport coefficients of produced matter.  It brings information on the pressure and stiffness of the equation of state of matter created during the earliest collision stages. The big surprise at RHIC  was the finding that $v_{2}$ increases by 50\% from the  top SPS energy $\sqrt{s_{NN}}$ =17.2 GeV to the top RHIC energy $\sqrt{s_{_{ \mathrm{NN}}}}$ =200 GeV \cite{qgp}.  The large value of $v_{2}$  observed at RHIC and later on at LHC \cite{ALICEflow} is one of the cornerstones of the perfect liquid bulk matter dynamics.  

For nucleons  $v_{1}$ reflects the hydrodynamic pressure that is created as the two projectile nuclei compress the interaction zone. A reduction of $v_{1}(y)$ slope $dv_{1}/dy$ in the vicinity of mid-rapidity has long been proposed as evidence of the softening on the equation of state associated with a phase transition. AGS data on protons show a monotonic decrease of $v_{1}(y)$ slope  from $\sqrt{s_{_{ \mathrm{NN}}}}$ =2.7 to 4.7 GeV \cite{AGS_flow}. STAR BES result (left panel of Fig.~\ref{fDiff_v2_sNN_muB})  shows that proton $v_{1}(y)$ slope changes sign from positive to negative between 7.6 and 11.5 GeV and remains small but negative up to 200 GeV.  In contrast to that $v_{1}(y)$ slope for net-protons, which is a proxy for the directed flow contribution from baryon number transported to mid-rapidity,  shows a compelling non-monotonic behaviour through the BES region.  Observed double sign change can not be explained by the UrQMD transport model but is qualitatively consistent with a prediction of hydrodynamic model with a first order phase transition \cite{Stocker}. 

\begin{figure}[hbt] 
 \includegraphics[width=.36\linewidth,height =.28\linewidth]{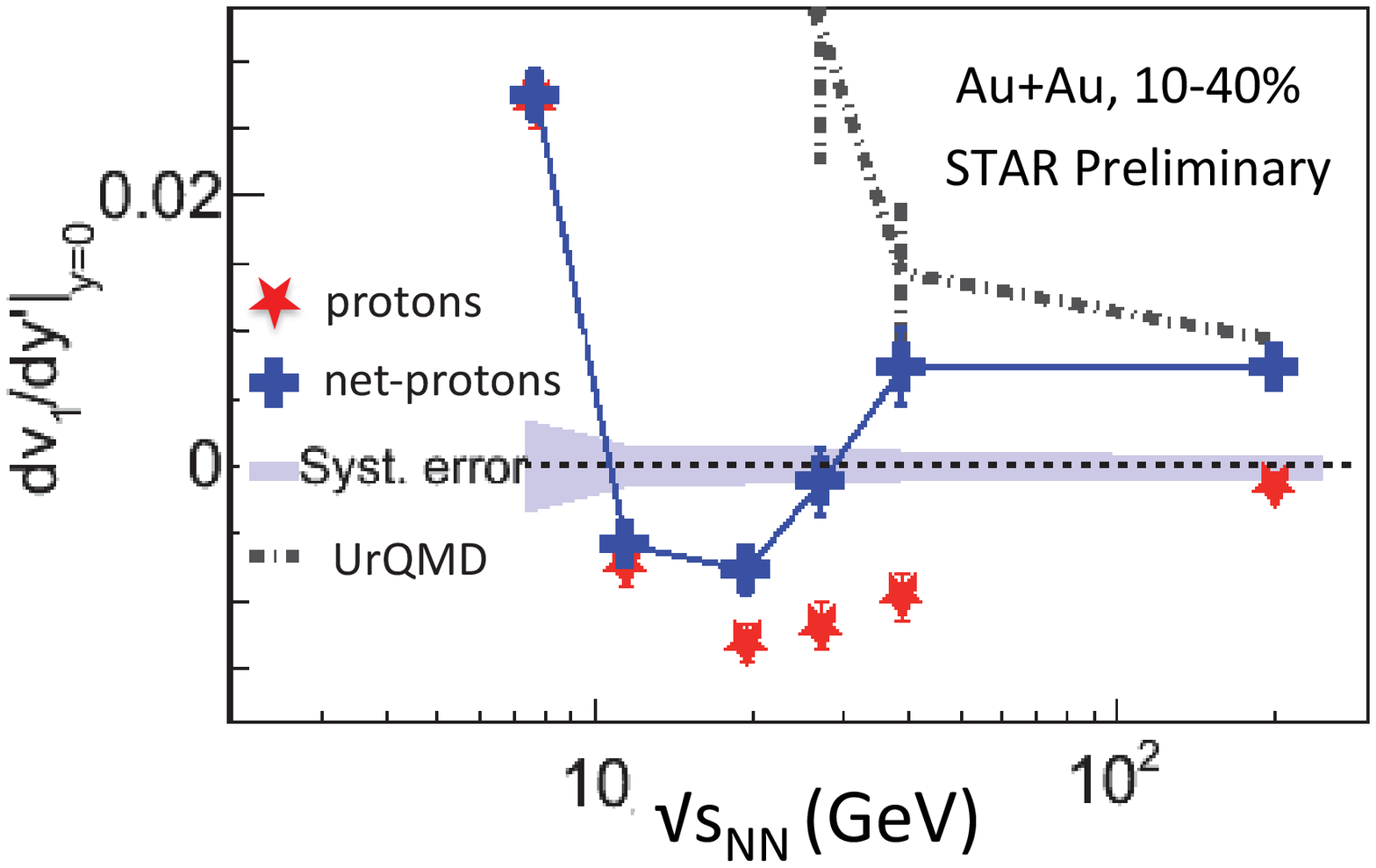} \hskip -.2cm
 \includegraphics[width=.335\linewidth]{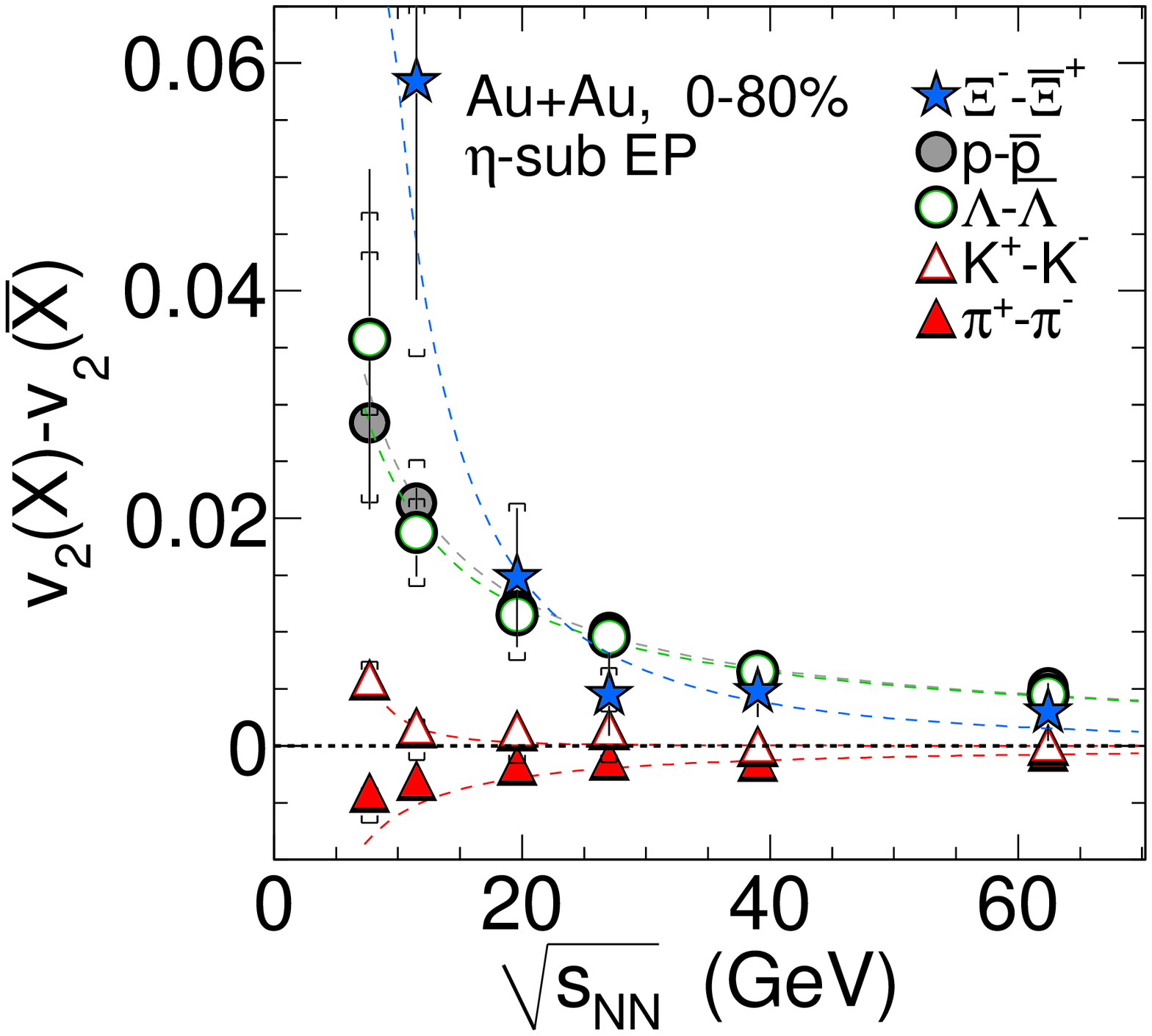} \hskip -.5cm
 \includegraphics[width=.335\linewidth]{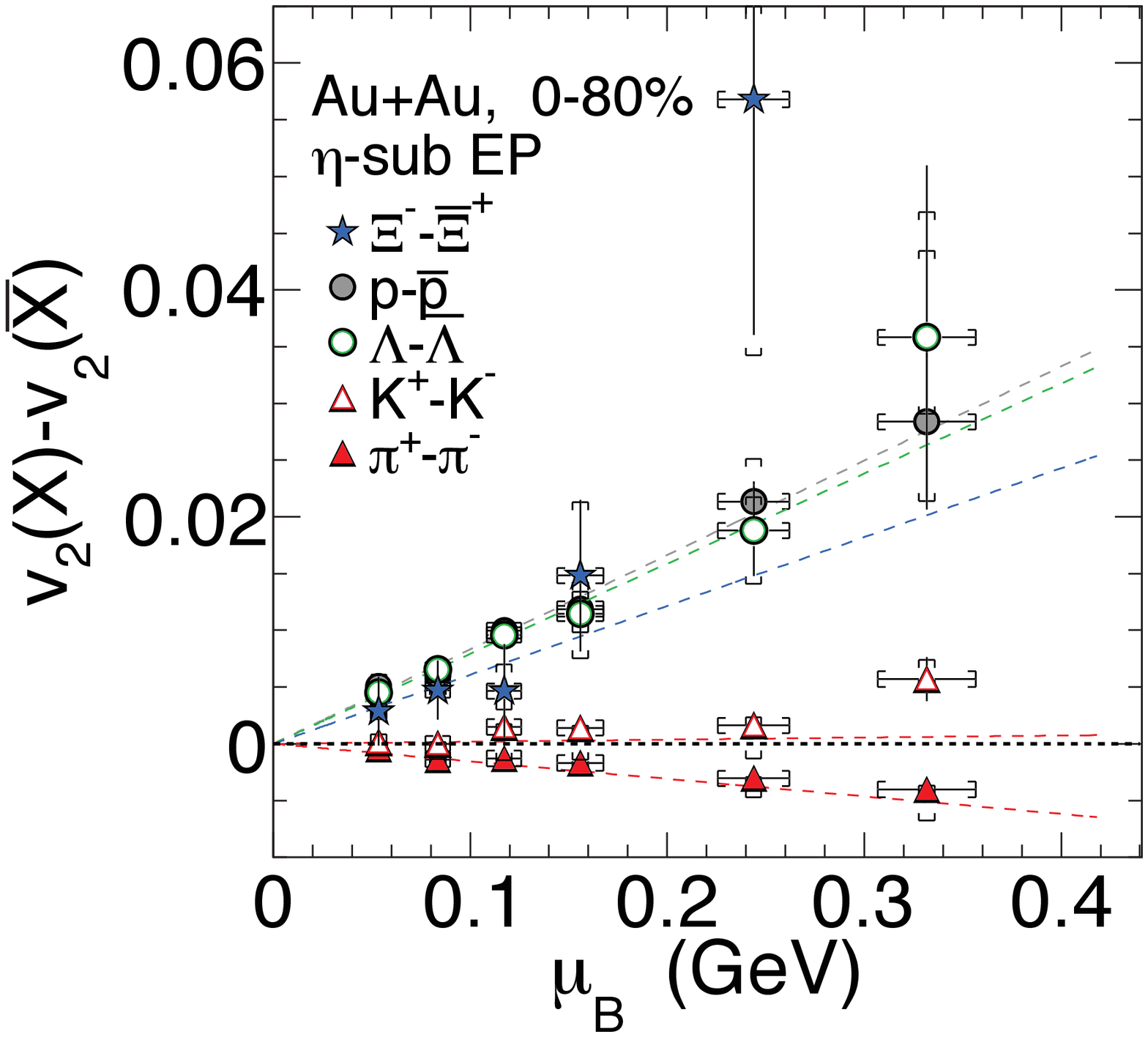}
  \caption{(Color online) Left panel: Slope of the directed  flow $v_{1}(y)$ near mid-rapidity of protons (red stars) and net-protons (blue crosses) for 10--40\% central Au+Au collisions  as a function of $\sqrt{s_{NN}}$.  Middle and right panel: The difference in the $v_{2}$ values between a particle $X$ and its corresponding anti-particle ${\rm \overline{X}}$ (see legend) as a function of $\sqrt{s_{NN}}$ (middle, \cite{v2_diff_PRL}) and $\mu_{B}$ (right,  \cite{v2_antiparticle}) for 0--80\% central Au+Au collisions. The dashed lines in middle plot are fits using $f_{\Delta v_{2}}(\sqrt{s_{NN}}) = a \cdot s_{NN}^{-b/2}$, and lines through the origin are shown for the right plot.} 
 \label{fDiff_v2_sNN_muB}
\end{figure}

%Moreover,  the differential $v_{2}(p_{T})$, that characterizes the hydrodynamic response to the initial geometry,  seems to be unchanged between the top RHIC energy and LHC energy of $\sqrt{s_{NN}}$=2.76 TeV~\cite{ALICEflow}.  
STAR BES results on $v_{2}$ of identified hadrons in the regime where the relative contribution of baryon and mesons vary significantly are presented in \cite{v2_antiparticle}.  Of particular interest is the energy dependence of the $v_{2}$ difference between particles $X$ ($p$, $\Lambda$, $\Xi^{-}$, $\pi^{+}$, $K^{+}$) and anti-particles $\overline{X}$ ($\bar{p}$, $\overline{\Lambda}$, $\overline{\Xi}^{+}$, $\pi^{-}$, $K^{-}$) \cite{v2_diff_PRL, v2_antiparticle} .  This difference is denoted in the following as $v_{2}(X)-v_{2}(\overline{X})$ and is shown on middle panel of Figure~\ref{fDiff_v2_sNN_muB} as a function of the beam energy $\sqrt{s_{NN}}$.   At 62.4 GeV, the $v_{2}$ difference for mesons is close to zero, whereas the baryons show positive difference. The difference increases for all particle species as the energy decreases. The baryons show a steeper rise compared to the mesons. The pions and kaons show a similar trend, but opposite with respect to their charge. Also, the protons and $\Lambda$ are very similar at all energies. Compared to the protons and $\Lambda$, the $\Xi$ show a slightly smaller difference at higher energies, but a larger difference at lower energies. 
%The difference in $v_{2}(\sqrt{s_{NN}})$ shown in the middle panel of Fig.~\ref{fDiff_v2_sNN_muB}  was parametrized with $f_{\Delta v_{2}}(\sqrt{s_{NN}}) = a \cdot s_{NN}^{-b/2}$. The fit results of the parameters {\it a} and {\it b} can be found in \cite{v2_antiparticle}.  
In right panel of Fig.~\ref{fDiff_v2_sNN_muB}, the $v_{2}$ difference is shown as a function of the baryonic chemical potential, $\mu_{B}$ \cite{v2_antiparticle}. A parametrization from \cite{Tiwari} was used to determine the $\mu_{B}$ values for each beam energy.  A linear increase of the $v_{2}$ difference with $\mu_{B}$ is observed for all particle species from 62.4 GeV down to 7.7 GeV. 
%Only at 11.5 GeV a ~2$\sigma$ deviation for the $\Xi$ and at 7.7 GeV a deviation for the kaons was found. 
This linear scaling behavior suggests that the baryon chemical potential is directly connected to the difference in $v_{2}$ between particles and anti-particles.

\section{The BES Phase-II and STAR in Fixed Target Mode}
To strengthen the message from Phase-I, higher statistics at lower energies  is needed, especially at 7.7 and 11.5 GeV. 
Statistics for several important observables, such as $\phi$-meson $v_2$ \cite{v2_antiparticle} or higher moments of net-protons distributions \cite{net-proton}, are not sufficient to draw quantitative conclusions. In order to confirm the trends between 11.5 and 19.6 GeV, 
another energy point is needed around 15 GeV in order to fill the 100 MeV gap in $\mu_B$.
This will be part of the BES Phase-II program proposed by  STAR. Simulation results indicate that with electron cooling, the luminosity could be increased by a factor of about 3--5 at 7.7 GeV and about 10 around 20 GeV~\cite{ecool}.   An additional improvement in luminosity may be possible by operating with longer bunches at the space-charge limit in the collider~\cite{longbunch}. Altogether  a factor or 10 improvement in luminosity is expected after these modifications. This will not only allow the precision measurements of the above important observables but will also help in the measurements of rare probes such as dileptons \cite{Geurts} and hypertritons \cite{hypertriton}. 

To maximize the use of collisions provided by RHIC for the BES program, an option to run STAR as a fixed-target experiment
is under consideration. A fixed Au target is to be installed in the beam pipe.  This will allow the extension of the $\mu_B$ range from 400 MeV to about 800 MeV covering  thus a substantial portion of the phase diagram.  The data taking can be done concurrently during the normal RHIC running. This proposal will not affect the normal RHIC operations. 

%\vspace{-.3cm}
\section{Summary}
Recent results from Phase-I of the RHIC BES program have substantially extended our knowledge of hot and dense de-confined QCD matter.  Interesting but smooth patterns of energy dependence are seen in most of the presented  analyses.  A linear increase of the $v_{2}$ difference with $\mu_{B}$ is observed for all particle species from 62.4 GeV down to 7.7 GeV. 
Energy dependence of $v_{1}$ of net-protons  shows a compelling non-monotonic behaviour through the BES region.

%\vspace{-.3cm}
\section*{Acknowledgements}
The work has been supported by the grant 13-02841S of the Czech Science Foundation.

%\vspace{-.3cm}

\end{document}